\input epsf.sty
\newdimen\psfigsize
\def\psfigure#1 #2 #3 #4 #5{
    \begin{figure}[tbh]
      \vbox{
        \null\vskip-0.1in\hskip#2
        \epsfxsize=#1
        \epsfbox{#4}
        \vskip -0.1in
        \caption {#5 \label{#3}}
        \vskip 0.0truein plus0.2truein
      }
    \end{figure}
}

\def\nabstar#1{\nabla\kern-0.5pt\smash{\raise 4.5pt\hbox{$\ast$}}
               \kern-4.5pt_{#1}}

\def\drvstar#1{\partial\kern-0.5pt\smash{\raise 4.5pt\hbox{$\ast$}}
               \kern-5.0pt_{#1}}

\def\newline{\relax\ifhmode\null\hfil\break\else\nonhmodeerr@\newline\fi}
\def\frac#1#2{{#1\over#2}}
\def\text#1{{\hbox{\rm #1}}}

\newcommand{\beq}{\begin{equation}}
\newcommand{\eeq}{\end{equation}}
\newcommand{\bea}{\begin{eqnarray}}
\newcommand{\eea}{\end{eqnarray}}
\def\Id{ \mbox{1\hspace{-1.2mm}I} }
\def\BE{\begin{equation}}
\def\EE{\end{equation}}
\def\BA{\begin{eqnarray}}
\def\EA{\end{eqnarray}}
\def\BAN{\begin{eqnarray*}}
\def\EAN{\end{eqnarray*}}
\def\LL{\left}
\def\RR{\right}
\def\nn{\nonumber\\}

\def\tr{\mathrm{tr}}

\def\bpsi{\bar\psi}
\def\gm5{\gamma_5}

\def\bpsi{\bar{\psi}}

\def\CT{{\cal T}}
\def\anx{{\cal A}_n(x)}

\def\BE{\begin{equation}}
\def\EE{\end{equation}}
\def\BA{\begin{eqnarray}}
\def\EA{\end{eqnarray}}
\def\BAN{\begin{eqnarray*}}
\def\EAN{\end{eqnarray*}}
\def\LL{\left}
\def\RR{\right}
\def\nn{\nonumber\\}

\def\tr{\mbox{tr}}

\def\gm5{\gamma^5}

\def\bpsi{\bar{\psi}}

\def\text#1{{\rm #1}}
\documentstyle[twoside,fleqn,espcrc2]{article}
\begin{document}
\title{Chiral Anomaly and Index Theorem on a finite lattice}
\author{Ting-Wai Chiu%
\address{Department of Physics, National Taiwan University,
Taipei, Taiwan 106, Republic of China.}
\thanks{%
This work was supported by the National Science Council, R.O.C. under the
grant numbers NSC88-2112-M002-016 and NSC89-2112-M002-017}}

\begin{abstract}

The condition for a lattice Dirac operator $ D $ to reproduce correct
chiral anomaly at each site of a finite lattice for smooth background
gauge fields is that $ D $ possesses exact zero modes satisfying the
Atiyah-Singer index theorem. This is also the necessary condition for
$ D $ to have correct fermion determinant ( ratio ) which plays the
important role of incorporating dynamical fermions
in the functional integral.

\end{abstract}

\maketitle

\section{INTRODUCTION}

Chiral anomaly plays an important role in particle physics.
In QCD, the chiral anomaly breaks the global chiral symmetry of
massless QCD through the internal quark loop.
When the axial current couples to two external photons,
the axial anomaly \cite{ABJ} provides a proper account
of the decay rate of $ \pi^{0} \to \gamma \gamma $.
For the flavor singlet axial current coupling to two gluons,
the presence of chiral anomaly but no associated Goldstone boson
posed the $ U_A(1) $  problem which was resolved by 't Hooft,
after taking into account of the topologically nontrivial gauge
configurations, the instantons \cite{tHooft76}.
This may also provide an explanation why the
$\eta$ ( $\eta'$ ) particle is much heavier than the $\pi$'s ( $K$'s ).

Lattice QCD should be designed to provide nonperturbative and
quantitive answers to all these and related problems.
However, these goals could not be attained if the
lattice Dirac fermion operator does not reproduce correct chiral
anomaly or the fermion determinant ( ratio ) on a finite
lattice {\em without fine-tuning any parameters.}

A basic requirement for lattice Dirac operator $ D $ is that, on a
finite lattice with any prescribed smooth background gauge field
which has integer topological charge, $ D $ possesses {\em exact}
zero modes satisfying the Atiyah-Singer index theorem
( i.e., $ D $ is {\em topologically proper} )
without fine tuning any parameters.

For any topologically proper $ D $, the sum of the anomaly function
over all sites must be correct, and is equal to two times of the
topological charge of the background gauge field.
If the anomaly function of $ D $ at some of the
sites do not agree with the Chern-Pontryagin density, we can perform
the following {\em topologically invariant} transformation \cite{twc98:6a}
\bea
\label{eq:twc}
\CT(R) : \hspace{4mm} D'= \CT(R)[D] \equiv D ( \Id + R D )^{-1}
\eea
with some operator $ R $ such that $ D' $ is local, then the anomaly
function of $ D' $ would be in good agreement with the Chern-Pontryagin
density at each site\footnote{ Here we have assumed that the size of
the finite lattice is large enough such that the finite size effects
can be neglected, i.e., the size of the lattice is much larger than
the localization length of $ D' $.}.
It suffices to choose $ R $ to be local in the
position space and trivial in the Dirac space.
Note that it is not necessary to perform any fine tunings of $ R $
since any local $ D' $ would give the correct chiral anomaly.
The set of transformations, $ \{ \CT(R) \} $,
form an abelian group with the group parameter space $\{ R \}$ \cite{twc99:6}.
From the physical point of view, we must require the existence of
a chirally symmetric $ D_c  = \CT(R_c) [D] $ such that in the free
fermion limit $ D_c(p) \to i \gamma_\mu p_\mu $ as $ p \to 0 $,
otherwise, $ D $ is irrelevant to massless QCD.
In general, we assume that there exists
\BAN
\label{eq:DcRc}
D_c = D ( \Id + R_c D )^{-1}
\EAN
such that
$ D_c \gamma_5 + \gamma_5 D_c = 0 $.
Then one obtains
\BA
\label{eq:gwr}
D \gamma_5 + \gamma_5 D = - 2 D R_c \gamma_5 D
\EA
This is the rejuvenated Ginsparg-Wilson relation \cite{gwr}.
In general, for any $ D $, if the chirally
symmetric $ D_c = \CT(R_c)[D] $ exists, then $ D $
must satisfy the Ginsparg-Wilson relation.
Thus the Ginsparg-Wilson relation does {\em not}
specify the topological characteristics of a Dirac operator.
Only a {\em topologically proper} $ D $ ( or its
transform $ D' = D ( \Id + R D )^{-1} $ ) is suitable
for lattice QCD.

\section{THE ANOMALY FUNCTION}

Consider a lattice Dirac operator $ D $
which breaks the usual chiral symmetry according to
$ D \gamma_5 + \gamma_5 D = B $,
where $ B $ is a generic irrelevant operator, and the lattice is
finite with periodic boundary conditions.
Then we can obtain \cite{twc99:6}
\BAN
\partial^{\mu} J^5_{\mu}(x) =
    \bpsi_x \gm5 ( D \psi )_x  + (\bpsi D)_x \gm5 \psi_x     \\
     - \frac{1}{2} ( \bpsi B )_x  \psi_x
     - \frac{1}{2} \bpsi_x  ( B \psi )_x
\EAN
which satisfies the conservation law,
\BA
\label{eq:Q5}
\sum_x \partial^{\mu} J^5_{\mu}(x) = 0.
\EA

If $ D $ {\em does not possess exact zero modes} in the background gauge
field, then the fermionic average of $ \partial^{\mu} J^5_{\mu}(x) $
is
\BAN
\LL<\partial^{\mu} J^5_{\mu}(x)\RR> =
\frac{1}{2} \ \tr \left[(B D^{-1})(x,x) + (D^{-1} B)(x,x) \right]
\EAN
\BAN
\equiv {\cal A}_n (x) \ \ \mbox{ : anomaly function}
\EAN
However, for any $ D $ which does not possess exact zero modes
in the background gauge field, the sum of the anomaly function
over all sites must vanish due to the conservation law,
\BAN
\sum_x \LL< \partial^{\mu} J^5_{\mu}(x) \RR> = \sum_x \anx = 0
\EAN
This implies that if $ \anx $ is not zero identically for all $ x $,
then it must fluctuate from positive to negative values with respect
to $ x $. The latter case is exactly what happens to the anomaly function
of the standard Wilson-Dirac fermion operator.

On the other hand, if $ D $ {\em possesses exact zero modes} in
topologically nontrivial background gauge fields, then $ D^{-1} $
is not well defined. In this case, one introduces an infinitesimal
mass $ m $ and replace $ D $ by
$ \hat{D} = D + m f[D] $ ( where $ f[D] $ is any functional of $ D $,
which has eigenvalue one for the exact zero modes of $ D $ ),
and finally take the limit ( $ m \to 0 $ ), then we obtain
\BA
\LL< \partial^{\mu} J^5_{\mu}(x) \RR> &=& \anx
  + 2 \sum_{s=1}^{N_{+}} [\phi_s^{+}(x)]^{\dagger} \phi_s^{+} (x) \nn
  && - 2 \sum_{t=1}^{N_{-}} [\phi_t^{-}(x)]^{\dagger} \phi_t^{-} (x)
\label{eq:divJ5_z}
\EA
where $ \phi_s^{+} $ and $ \phi_t^{-} $ are normalized eigenfunctions
of $ D $ with eigenvalues $ \lambda_s = \lambda_t = 0 $ and
chiralities $ +1 $ and $ -1 $ respectively\footnote{ Here we assume
that $ D $ is normal, $ D D^{\dagger} = D^{\dagger} D $, and $ D $
satisfies the hermiticity condition $ D^{\dagger} = \gamma_5 D \gamma_5 $. },
and the anomaly function is
\BAN
\anx = \lim_{m \to 0 } \frac{1}{2} \ \tr
                  \left[  ( B \hat{D}^{-1} ) (x,x)
                         +( \hat{D}^{-1} B ) (x,x) \right]
\EAN

\section{THE INDEX THEOREM}

Summing Eq. (\ref{eq:divJ5_z}) over all sites
and using the conservation law Eq. (\ref{eq:Q5}), we obtain
\BA
\label{eq:index_thm}
N_{-} - N_{+} = \frac{1}{2} \sum_x  \anx
\EA
This is the index theorem for lattice Dirac operator on a
finite lattice, in {\em any} background gauge field.
However, it does {\em not} necessarily imply the existence of exact
zero modes nor the compliance with the Atiyah-Singer index theorem.

Since $ D $ does not have exact zero modes in a trivial
gauge background, the index of $ D $ must be proportional
to $ Q $ of the smooth background gauge field.
If $ Q $ is an integer, then the proportional constant must be
an integer, otherwise their product in general cannot be an integer.
Denoting this integer multiplier by $ c[D] $, we have
\BA
\label{eq:cD}
N_{-} - N_{+} = c[D] \ Q \ n_f
\EA
where $ n_f $ is number of fermion flavors.
Here we have assumed that $ c[D] $ is constant for smooth background
gauge fields. This is a reasonable assumption since $ c[D] $ is an intrinsic
characteristics of $ D $. However, when the gauge field becomes rough,
we expect that the index theorem with integer $ c[D] $ would break down.
If one insists that it holds even for {\em rough} gauge configurations,
then $ c[D] $ cannot be an integer constant due to the highly nonlinear
effects of the gauge field.
In general, $c[D]$ is a {\em rational number} functional of $ D $,
which in turn depends on the gauge configuration, but it becomes an integer
constant only for smooth gauge configurations.
The {\em topological characteristics} of $ D $, $ c[D] $,
was first discussed in ref. \cite{twc98:9a}
and was investigated in ref. \cite{twc98:10a}
for the Neuberger operator \cite{hn97:7}.


\section{A GENERAL SOLUTION FOR $\anx$ }

Now we try to obtain a general solution of $ \anx $ satisfying
(\ref{eq:index_thm}) and (\ref{eq:cD}), i.e.,
\BAN
\frac{1}{2} \sum_x \anx = N_{-} - N_{+} = c[D] \ Q \ n_f.
\EAN
Consider the gauge configuration with constant field tensors,
\BAN
F^{a}_{12}(x) =  \frac{ 2 \pi q_1 }{ L_1 L_2 }, \ \
F^{a}_{34}(x) =  \frac{ 2 \pi q_2 }{ L_3 L_4 }
\EAN
and other $F$'s are zero, where $ q_1 $ and $ q_2 $ are integers.
Then the topological charge of this configuration is
\BAN
\label{eq:Q_top}
Q &=& \frac{ \tr \{ t_a t_b \}}{32 \pi^2}
      \sum_x \epsilon_{\mu\nu\lambda\sigma}
     F_{\mu\nu}^{a}(x) F_{\lambda\sigma}^{b}(x+\hat\mu+\hat\nu)
         \\
  &=& n \ q_1 \ q_2
\EAN
which is an integer.
If $ D $ is local, then $ \anx $ is constant for all $ x $.
We immediately obtain
\BAN
\hspace{-2mm} && \anx \\
&=& \frac{n_f}{16 \pi^2} c[D] \epsilon_{\mu\nu\lambda\sigma}
              F_{\mu\nu}^{a}(x) F_{\lambda\sigma}^{b}(x+\hat\mu+\hat\nu)
              \tr \{ t^a t^b \}
\EAN
This is the particular solution of $ \anx $.
The complementary solution to the homogeneous equation
$ \sum_x \anx = 0 $ in general can be written as
\BAN
{\cal A}^{(h)} (x) = \sum_\mu \ [ G_\mu(x) - G_\mu(x-\hat\mu) ] \equiv
\partial_\mu G_\mu (x)
\EAN
where $ G_\mu(x) $ is any local and gauge invariant function.
Then the {\em general solution of the anomaly function} is the sum of the
complementary solution and the particular solution,
\BA
\nn
&&  \anx   \nn
&=& \frac{n_f}{16 \pi^2} c[D] \epsilon_{\mu\nu\lambda\sigma}
     F_{\mu\nu}^{a}(x) F_{\lambda\sigma}^{b}(x+\hat\mu+\hat\nu)
     \tr \{ t^a t^b \} \nn
&& + \sum_\mu \ [ G_\mu(x) - G_\mu(x-\hat\mu) ]
\label{eq:anx}
\EA
In general, $ G_\mu (x) $ depends on the background gauge field
and $ D $.
Only when $ \partial_\mu G_\mu (x) $ vanishes identically for any
smooth gauge background, $ D $ can reproduce
the correct chiral anomaly provided $ c[D] = 1 $.
( e.g., the Neuberger operator \cite{hn97:7} ).
On the other hand, the pathologies of the
standard Wilson-Dirac operator are $ c[D] = 0 $ and
$ {\cal A}^{(h)} (x) \ne 0 $.

Now we consider $ D $ with $ \partial_\mu G_\mu (x) = 0 $
in a constant gauge background.
Then we introduce local fluctuations to the background
gauge field and/or increase its topological charge $ Q $.
Then $ c[D] $ may remain constant provided
that the local fluctuations of the gauge fields are not too violent and/or
$ Q $ is not too large. When the gauge background becomes so rough
that $ D $ is non-local, then $ \partial_\mu G_\mu (x) \ne 0 $
but the index theorem may still hold with the same $ c[D] $.
If we keep on increasing the roughness of the gauge background,
then $ D $ will undergo a topological phase transition, and $ c[D] $
will become another integer or even a fraction \cite{twc99:6}.

\end{document}